# Performance Evaluation of VoLTE Based on Field Measurement Data

Ayman Elnashar, Mohamed A. El-Saidny, and Mohamed Yehia

**Abstract**— Voice over Long-Term Evolution (VoLTE) has been witnessing a rapid deployment by network carriers worldwide. During the phases of VoLTE deployments, carriers would typically face challenges in understanding the factors affecting the VoLTE performance and then optimizing it to meet or exceed the performance of the legacy circuit switched (CS) network (i.e., 2G/3G). The main challenge of VoLTE service quality is the LTE network optimization and the performance aspects of the service in different LTE deployment scenarios. In this paper, we present a detailed practical performance analysis of VoLTE based on commercially deployed 3GPP Release-10 LTE networks. The analysis evaluates VoLTE performance in terms of real-time transport protocol (RTP) error rate, RTP jitter and delays, block error rate (BLER) in different radio conditions and VoLTE voice quality in terms of mean opinion score (MOS). In addition, the paper evaluates key VoLTE features such as RObust Header Compression (ROHC) and transmission time interval (TTI) bundling. This paper provides guidelines for best practices of VoLTE deployment as well as practical performance evaluation based on field measurement data from commercial LTE networks.

## 1 Introduction

Voice over Long-Term Evolution (VoLTE) is an IP multimedia system (IMS)-based voice service over the LTE network [1]. The IMS supports various access and multimedia services and has recently become the standard architecture of evolved packet core (EPC) [1], [2]. 3GPP has adopted GSMA IR.92 IMS profile for voice and SMS [3] and GSMA IR.94 IMS profile for conversational video [4] to provide high quality IMS-based telephony services over LTE radio access network. The profiles define optimal sets of existing 3GPP-specified functionalities that network infra-vendors, service providers and handset manufacturers can use to offer compatible LTE voice/video solutions. Therefore, the commercial deployment of VoLTE mandates extensive testing between terminals and networks including LTE radio access network (i.e., eNodeB (eNB)), LTE EPC, and IMS. In addition to these challenges, the VoLTE optimization for different radio/loading conditions to find an acceptable tradeoff between the user's experience and network deployment complexities has led to a substantial delay in widely adopting VoLTE service. In this paper, we address these challenges by providing the best practices for VoLTE related features and practical performance evaluation based on field-testing results from commercial LTE 3GPP Rel 10 networks.

The deployment of VoLTE brings variety of benefits to telecom operators as voice is still the main source of revenue. Hence, telecom carriers need voice evolution to effectively compete with over-the-top (OTT) voice over IP (VoIP) applications that create a significant load on the mobile broadband networks and accordingly affecting other services. VoLTE also improves the spectral efficiency and reduces network costs compared to legacy circuit switched (CS) networks. Moreover, the spectral efficiency of LTE networks is higher than the traditional GSM/UMTS networks, which makes VoLTE a really suitable voice solution in 4G networks [5], [6]. For the same channel bandwidth, an LTE cell offers twofold cell-edge throughput of a UMTS cell [5], [6].

VoLTE enhances the end-user experience by providing a better quality of experience (QoE). VoLTE is equipped with a numerous set of integrated features that improves QoE aspects such as better call setup time, higher efficiency in deep coverage conditions and lower battery consumption. Moreover, VoLTE supports different range of speech codec rates i.e., adaptive multi-rate (AMR) with both wide-band (AMR-WB) and narrow-band (AMR-NB) and enhanced voice services (EVS) codec. While high definition (HD) voice is being rolled out by telecom operators using AMR-WB with audio bandwidth of up to 7 kHz, 3GPP Rel-12 introduced EVS codec to offer up to 20 kHz audio bandwidth [7]. EVS at 13.2kbps provides super-wideband (SWB) voice quality at comparable bit-rate to AMR and AMR-WB and offers high robustness to jitter and packet losses.

The feasible solutions for providing voice call and service continuity over LTE-based networks, a comparison between various aspects of these solutions, and a possible roadmap that mobile operators can adopt to provide seamless voice over LTE are provided in [8]. A comprehensive evaluation and validation of VoLTE quality of service (QoS) is provided in [9]. The results in [9] give clear evidence that the VoLTE service fulfills the ITU-R and 3GPP standard requirements in terms of end-to-end delay, jitter and packet loss rate. The VoLTE performance in terms of Quality of Service (QoS) is evaluated and validated using OPNET modeler wireless suite in [9]. The railway voice communication based on VoLTE is proposed in [10]. The simulation results in [10] indicate that VoLTE is a viable option for providing railway voice communication i.e., GSM-R. Therefore, LTE is proven to be a strong candidate for becoming the future communication network for railways [11]. The feasibility of semi-persistent scheduling (SPS) for VoIP is analyzed in [12], which evaluates its performance in terms of throughput of random access delays and traffic channels. A methodology to evaluate the Voice-over-IP (VoIP) capacity and performance of orthogonal frequency-division multiple-access (OFDMA)-based systems is provided in [13]. This methodology can also be applied to VoLTE. A method for estimating cell capacity from network measurements in a multiservice long-term evolution (LTE) system is presented in [14].

The performance of circuit switched fallback (CSFB), VoLTE and enhanced single radio voice call continuity (eSRVCC) are key elements to guarantee seamless voice experience within the LTE/UMTS/2G networks [15]-[20], especially that a mix of VoLTE and CSFB devices can coexist in the same LTE network. The three features were analyzed in [6], [20] to evaluate the call setup delays and the details behind the interruption time within LTE network or over eSRVCC. It is demonstrated in [20] that VoLTE provides a better end-user experience in terms of call setup delay. However, the eSRVCC voice interruption is higher than the LTE intra-frequency interruption by ~ 200 msec, but still within the acceptable audio quality range of 300 msec. A comprehensive performance evaluation of RObust Header compression (ROHC) for VoLTE by means of a testbed implementation is presented [21]. In [22], it is envisaged that quality of call measured through mean opinion score (MOS) is always better in LTE compared with UMTS network. The VoLTE MOS condition only becomes unstable upon reaching very poor RF condition – where reference signal received power (RSRP) < -117dBm and reference signal received quality (RSRQ) < -12dB. The implementation of transmission time interval (TTI) bundling feature helps to improve the uplink (UL) coverage and minimizes the BLER. It is proposed in [22] that serving cell SRVCC threshold RSRP = -108dBm for networks without TTI bundling and RSRP = -117dBm for networks with TTI bundling deployment. However, this paper demonstrates that RSRP = -110 dBm is the optimum threshold to maintain voice quality even with TTI bundling deployment, as real-time transport protocol (RTP) performance can be adversely impacted beyond this level.

The SRVCC requires UE to support the ability to transmit and receive, simultaneously, on both networks (Packet Switch (PS) such as LTE and Circuit Switch (CS) such as 3G). SRVCC went through different stages in the standard in order to reduce the voice interruption time that impacts the user's experience as well as improving the call setup success rate at different stages of the VoLTE call. 3GPP has started with the support of SRVCC in Release 8/9 and then enhanced the mechanism to support enhanced SRVCC (eSRVCC) in Release 10 [20]. The main target of the eSRVCC is to reduce the voice interruption during the inter-technology handover. eSRVCC targets an interruption of < 300 msec.

In this paper, we present comprehensive practical performance analysis of VoLTE performance based on commercially deployed 3GPP Release-10 LTE networks. The analysis demonstrates VoLTE performance evaluation in terms of RTP error rate, RTP jitter and delays, BLER and VoLTE voice quality in terms of MOS. In addition, this paper evaluates key VoLTE features such as ROHC, TTI bundling and SPS. The remaining of the paper are organized as follows. VoLTE principles are summarized in section II. Main VoLTE features along with deployment best practices are outlined in section III. Testing environment is explained in section IV. Practical performance analysis is provided in section V. The conclusions and future work are summarized in section VI.

## 2 VoLTE Principles

The IMS server carries all sources of VoLTE traffic and provides functions including subscribers' registration, authentication, control, routing, switching, media negotiation and conversion. The voice codec type and rate of VoLTE service are negotiated at call setup by the UE (User Equipment) and IMS. The eNB provides only the media-plane bearer channel and the IMS signaling is transparent to the air interface.

After the VoLTE call is established and the media packets start flowing, the eNB performs dynamic scheduling and uses power control policies that are suitable for such scheduling. The eNB selects a modulation and coding scheme (MCS) index and physical resource blocks (PRBs) for voice users similar to the mechanism in scheduling the PS data. The main purpose of the VoLTE scheduling technique is to maintain continuous transmission on uplink and downlink in a way that minimizes the packet delays. The data packets for voice services have a relatively small and fixed size, and therefore, scheduling RTP packets on the radio and core network requires stringent delay budget to control the inter-arrival time that minimizes the jitter.

There are several protocol interfaces used between the UE's IMS client and network's IMS server. They are the language between the IMS client/server to exchange either signaling information or actual media packets (i.e. voice packets). The summary of these protocols is as follows:
a) SIP: delivers IMS signaling to negotiate a media session between two users.
b) RTP: delivers IMS media packets.
c) RTP control protocol (RTCP): used to synchronize streams by providing feedback on QoS information.
d) IP Security (IPSec): used to carry the authentication and key agreement (AKA) as a secured tunnel for IMS clients.

Due to the delays in VoLTE network from different network elements (i.e., eNB, EPC, IMS, and transport network), the RTP packets inter-arrival time can vary in time. The RTP packets during talk spurts are generated every 20ms. However, the packets do not arrive precisely at that exact interval. This means that the VoLTE packets cannot be played out as they arrive due to variance in packet arrivals. Jitter is defined as a statistical variance of the RTP data packet inter-arrival time. In PS networks, the occurrence of variable delay is much higher than the values in CS networks. Figure 1 illustrates the concept of jitter and the delay requirements to keep the inter-arrival time within an acceptable range for better jitter buffer management. Jitter buffers are used to change asynchronous packet arrivals into a synchronous stream by tuning variable network delays into constant delays at the destination. The role of the jitter buffer is to set a trade-off between delay and the probability of interrupted playout because of late packets. Late or out-of-order packets are discarded. IMS clients shall implement adaptive jitter buffers to overcome these issues by dynamically tuning the jitter buffer size to the lowest acceptable value. Increasing the buffer size can increase latency. Even if the RTP packets remain in the correct sequence and there is zero packet loss, large variations in the end-to-end transmission time for the packets may cause degradation of audio quality that can only be fixed through the use of the jitter buffer. Typical RTP error rate is < 1% for audio and < 0.1% for video [23]. On the other hand, the radio network scheduler has a key role by securing sufficient scheduling resources in all radio conditions in order to keep the RTP error rate, jitter and end-to-end delays within a range that can be efficiently handled by the jitter buffer.

The VoLTE call includes an end-to-end voice/media flow transmitted on a dedicated guaranteed bit rate (GBR) bearer with QoS class identifier (QCI) = 1 through RTP protocol, user datagram protocol (UDP) or IP protocol. Another default bearer for session initiation protocol (SIP) signaling is established beforehand using QCI = 5, through UDP protocol, transmission control protocol (TCP) or IP protocol. The IMS network and evolved packet system (EPS) transfer the IMS access point name (APN) as well as the IMS PDN connection that host the two IMS bearers. The packet-switched (PS) services continue to use the default packet data network (PDN) connection with a non-GBR bearer (e.g. QCI = 9).

Radio bearers with QCIs of 1, 2, and 5 are established between two VoLTE UEs to carry conversational voice, signaling, and video, respectively. The eNB performs admission and congestion control for conversational voice (QCI = 1), signaling (QCI = 5), and video (QCI = 2). Moreover, the eNB performs dynamic or SPS scheduling and uses power control policies that are suitable for dynamic or SPS scheduling.

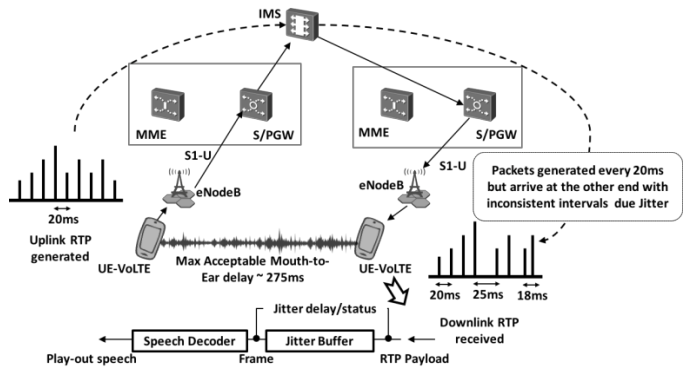

Fig. 1. The concept of jitter and jitter buffer management

## 3 Main VoLTE Features

The eNB scheduler is typically designed to efficiently schedule the UEs during the DL and UL transmissions for either small or large packet sizes. However, the VoLTE media stream has small voice packets with fixed inter-arrival intervals. Therefore, the eNB scheduler works with various special features specific for VoLTE in order to enhance the coverage, capacity and quality of the voice calls. These eNB features are ROHC, TTI Bundling, and SPS, which are all designed to assist the VoLTE in-call performance. These features will be presented in theses section.

### 3.1 Robust Header Compression

ROHC is a header compression protocol originally designed for real-time audio/video communication in wireless environment. As VoLTE media packets are transported as IP traffic, the generated headers of the IP protocols can be massively large. ROHC compresses the RTP,

UDP, and IP headers to reduce the size of the entire voice packets. This decreases the radio resource utilization on the cell-edge and therefore improves the overall cell coverage on both UL and DL. In addition, it reduces the number of voice packet segments, which improves the BLER associated with these smaller-size transmissions. This improves the VoLTE end-to-end delays and jitter.

The ROHC operation is illustrated in figure 2. The ROHC function in LTE is part of Layer-2 at the user plane of the UE and eNB. Both UE and eNB behave as a compressor and de-compressor for the user-plane packets on UL and DL. The compression efficiency depends on the ROHC operating mode and the variations in the dynamic part of the packet headers at the application layer. A header can be compressed to one byte with ROHC, which efficiently reduces the voice packet size. ROHC in LTE operates in three modes: U-Mode, O-Mode, and R-Mode (Unidirectional, Bidirectional Optimistic and Bidirectional Reliable, respectively). The reliability of these modes and overheads used for transmitting feedback are different. In U-Mode, packets can only be sent from the compressor to the de-compressor, with no mandatory feedback channel. Compared with O-Mode and R-Mode, U-Mode has the lowest reliability but requires the minimum overhead for feedback. In O-Mode, the de-compressor can send feedback to indicate failed decompression or successful context update. Therefore, it provides higher reliability than U-Mode but it generates less feedback compared to R-Mode. In R-Mode, context synchronization between the compressor and de-compressor are ensured only by the feedback. That is, the compressor sends the context updating packets repeatedly until acknowledgment is received from the de-compressor. Therefore, R-Mode provides the highest reliability but generates the maximum overhead due to the mandatory acknowledgment.

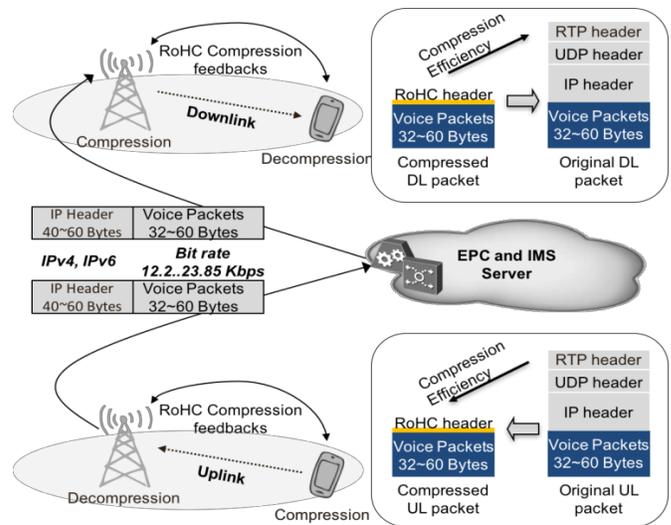

Fig. 2. ROHC Operation Mechanism

### 3.2 Transmission Time Interval Bundling

TTI bundling enables a data block to be transmitted in consecutive TTIs, which are packed together and treated as the same resource during the scheduling process. TTI bundling makes full use of hybrid automatic repeat request (HARQ) gains and therefore reduces the number of retransmissions and round trip time (RTT). When the user's channel quality is degraded or the transmit power is limited, the TTI bundling is triggered to improve the uplink coverage at cell-edge, reduce the number of different transmission segments at the radio link control (RLC) layer and the retransmission overhead. The main advantage of TTI bundling is enhancing VoLTE uplink coverage when the UE has limited uplink transmit power. Thus, it guarantees VoLTE QoS for cell-edge users. In a conventional scheduling mechanism, if the UE is not able to accumulate sufficient power to transmit a small amount of data, like a VoIP packet, the data packets can be segmented into smaller size packets that fit within the UE transmit power. Each segment will be transmitted with a separate HARQ process. However, this segmentation mechanism increases the amount of control information that needs to be transmitted. Therefore, the control channel load increases with the amount of segments since every segment requires new transmission resources on these channels. Additionally, the probability for HARQ feedback errors increases with the number of segments causing higher BLER at cell-edge. Therefore, the need to utilize better segmentation method like TTI Bundling is important.

For UEs at cell-edge RLC segmentation is done first, then TTI bundling transmit same packets four times in one scheduling period to extend the coverage by increasing uplink transmission reliability. The eNB decides when to activate or deactivate the TTI bundling for certain users based on the measured signal-to-interference noise ratio (SINR) and PRB usage on uplink. The data block in a bundle of TTIs, where the chunk of each bundle (up to 4 chunks), is modulated with different redundancy versions within the same HARQ identity. In the case of unsuccessful decoding of the HARQ identity, the eNB sends negative acknowledgement the UE, which requires retransmission. The resource allocation during this operation is restricted to a certain number of PRB and transport block size (TBS) in order to improve the probability of decoding at lower data rates. The mechanism to transmit same packets four times in one scheduling instance expands the coverage by increasing the uplink transmission reliability with a better success rate gain. In addition, it guarantees that VoLTE packets are transmitted at cell-edge when resources are limited to improve the latencies in bad radio conditions. TTI bundling is estimated to provide 2 - 4 dB uplink coverage improvement for VoLTE services which extend the cell radius for VoLTE service [25]. Figure 3 describes the TTI bundling mechanism and provide comparison with a scheduling mechanism that depends only on RLC segmentation procedures [26].

### 3.3 Semi-persistent scheduling

SPS reduces the control signaling overhead on the air interface while increasing the overall system capacity by means of scheduling the UE to receive the regular PRB resources in a fixed period with no scheduling grant on the physical downlink control channel (PDCCH). The feature is needed in scenarios where voice service UEs and other data UEs coexist in same cell. Thus, the increase in the number of PRBs allocated to voice service UEs will cause a decrease in the number of PRBs available to other data UEs, and consequently the cell throughput (capacity) will decrease. In SPS, the allocated traffic channel is released when a certain number of empty transmission slots are detected on the allocated data traffic channel. This is effectively achieved without changing the amount of resources or the packet size (i.e. MCS), at the beginning of SPS allocation period (e.g. 10, 20, 32…640 subframes). With this reduction mechanism in air interface, SPS scheduling offers up to 2.5x capacity improvement over the conventional dynamic scheduling in limited PDCCH scenario i.e., 5 PDCCH [27]. However, the dynamic scheduling mechanism is adopted in most deployed networks unless VoLTE load reaches the required SPS activation threshold (i.e. significant increase in load with sub-optimal control channel capacity). Therefore, this paper does not present practical results for SPS since it is not widely deployed.

## 4 TESTING Environment and VoLTE Main KPIs

In this paper, the VoLTE performance is assessed in terms of ROHC efficiency, RTP error rate, jitter, DL/UL BLER, handover delays and voice quality in terms of MOS. The data is processed from field measurement with a large sample size (i.e., long VoLTE calls during mobility) and results were averaged over two different LTE access networks (i.e., two different clusters) from two different suppliers and using two different smartphones to capture the main trend and mitigate any network or handset impact. The Two different network infra-vendors are tested where both are deployed with 1800MHz (20MHz channel) and collocated with UMTS at 2100MHz and GSM900/1800MHz. The Key Performance Indicators (KPIs) are estimated from the device side through post processing scripts to the collected logs from the device modem. Table I summarizes the LTE network parameters, average RF conditions in mobility and VoLTE related features. The testing is conducted in live commercial LTE networks with normal loading i.e., ~50%. The testing is conducted with and without PS data session to evaluate the impact of concurrent services (voice and data) on the VoLTE performance. The testing methodology is illustrated in figure 4. The eNB strategy is based on proportional fair scheduler with dynamic assignment based on load and channel conditions. QCI for VoLTE is always 1 and for data it can be 6, 8 or 9 (9 is used in the tested network). The scheduler differentiates voice and data based on QCI=1 or QCI = 9 giving priority to QCI=1 when a conflict occurs (e.g. channel condition is not suitable to schedule data from both services at same time, or in case of network congestion).

TABLE I
NETWORK AND VoLTE PARAMETERS

| Configuration | DL/UL |
|---|---|
| LTE System Bandwidth | 20 MHz |
| UE Category | 6 |
| MIMO Configuration | MIMO 2x2, TM3 |
| Mobility Speed | 80 km/h |
| **RF Conditions in Mobility** | **Average Values** |
| Serving Cell RSRP [dBm] | -83.8 |
| Serving Cell RSRQ [dB] | -8.7 |
| Serving Cell RSSI [dBm] | -54.9 |
| Serving Cell SINR [dB] | 20.2 |
| **VoLTE Relevant Parameter** | **Value** |
| ROHC | ON |
| TTI Bundling | ON |
| Dynamic Scheduling | ON |
| SPS | OFF |
| C-DRX for VoLTE configuration | ON (LongDrxCycle = 40 ms) |

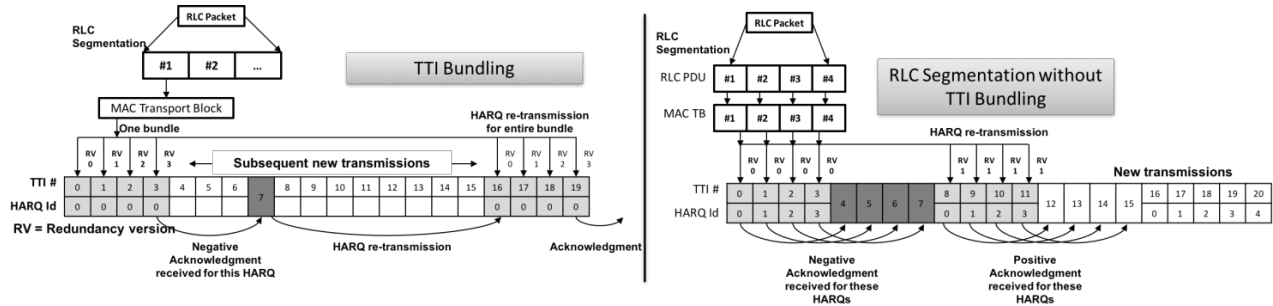

Fig. 3. TTI bundling and RLC segmentation procedures

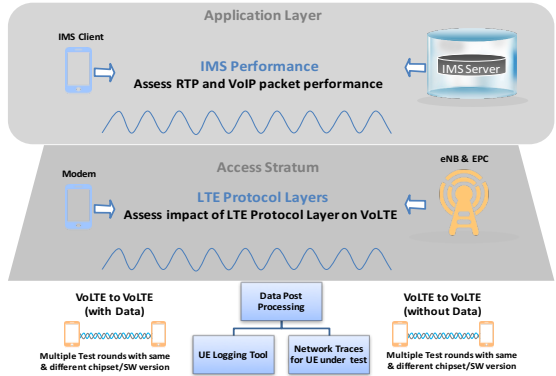

Fig. 4. Testing methodology summary

The focus of this paper is assessing the end-to-end performance of VoLTE calls. Therefore, we have conducted long VoLTE calls in the each cluster with a duration of ~2 hour. The main VoLTE KPIs demonstrated in this paper are summarized as follows:

Relative Jitter: inter-arrival time of subsequent RTP payloads calculated with reference to the previous RTP packet received in-sequence during a talk spurt. The jitter is calculated per RTP stream during talk spurts as follows:

$$Jitter = (IMS\ Time\ of\ current\ RTP - IMS\ Time\ of\ previous\ RTP) - (Time\ of\ current\ RTP - Time\ of\ previous\ RTP) \quad (1)$$

Then, the relative jitter $j(t)$ at time $t$ can be calculated as follows:

$$j(t) = abs((s(t) - s(t-1)) - (r(t) - r(t-1))), \quad (2)$$

*Where* abs (.) is the absolute value of a number and $s(t)$ is the RTP timestamp embedded inside the recent received RTP packet which is the actual timestamp of the RTP packet, $s(t-1)$ is the RTP timestamp embedded inside the previously received RTP packet which is the actual timestamp of the RTP packet, $r(t)$ is the timestamp of the recent received RTP packet i.e., the timestamp of arrival current RTP packet and $r(t-1)$: the timestamp of the previously received RTP packet i.e., the timestamp of the previous RTP packet.

<u>RTP DL Error Rate</u>: The percentage of the RTP packets that are not received by the UE based on RTP sequence number. The number of lost packets "$E$" is calculated per RTP flow by adding the number of RTP packets lost (i.e. the gaps in RTP sequence number). Similarly, the number of RTP packets successfully received "$N$" is calculated per RTP flow by counting the number of RTP sequence number and payload received in order. Then, the RTP downlink error rate is calculated as follows

$$RTP\ DL\ Error\ Rate = E/(E + N) \quad (3)$$

Both jitter and RTP error rate are calculated with reference to all RTP packets received over certain interval (e.g. 1 sec).

## 5 VoLTE Performance Evaluation

In this section, the VoLTE performance is evaluated in terms of ROHC, TTI, RTP error rate, jitter, BLER, handover delays (C-Plane and U-Plane) and voice quality in terms of MOS. All results are obtained from commercial networks as explained in the previous section.

### 5.1 ROHC Efficiency and Performance Evaluation

To evaluate the maximum capability of the ROHC, we have tested two scenarios; the first scenario for concurrent VoLTE and data connections and second scenario for a VoLTE standalone call. In both scenarios, the voice activity was continuous with minimum silent periods. In the first scenario, the packets at the radio side are typically multiplexing between both PS data and IMS within the same transmission time interval. However, because the compression takes place at the upper layers, then the impact on the actual size of the radio packet (efficiency) is not too much compared to the second scenario (IMS call only).

Table II provides practical results for ROHC header compression efficiency from real network deployment as described in section IV based on IPv4. The table illustrates DL/UL header size and average compression efficiency in mobility scenario for long VoLTE-to-VoLTE call covering 100 eNBs. As observed, the ROHC in both scenarios (i.e., with and without PS data) is capable of offering significant gain to the radio resources by reducing the packet size and compressing the headers with an average efficiency of 81% to 92% and overall average efficiency of 86.7%. Therefore, ROHC is very beneficial for VoLTE traffic transmitted alone or alongside other data traffic, which is a typical case in smartphone (i.e. background data is ongoing while the user is on a voice call).

In terms of channel rate saving, and using the practical values in table II, it is obvious that ROHC can reduce the transmission data rate on radio interface from:

$$Physical\ channel\ data\ rate = (P + H + O) * 8/I = 32.4\ kbps, \quad (4)$$

where P = AMR payload, H = average original header size, and O = other protocol headers, and I = RTP packets interval (i.e., 20ms), to

$$Physical\ channel\ data\ rate = (P + RH + O) * 8/I = 18.5\ kbps, \quad (5)$$

where RH is the average of compressed ROHC headers in all scenarios listed in table II which is 5.3 bytes. This indicates that ROHC boosts the air interface resources by almost twofold. This air interface saving provided by ROHC contributes directly to enhanced capacity and coverage. The exact capacity gain due to ROHC may vary depending on the deployment scenario (e.g. ROHC deployed with or without TTI Bundling features, ROHC mode of operations configured as explained in section II). It is expected that ROHC will even offer higher gain with IPv6 since the header is 60 bytes [24].

As evident from these results, the original voice packet sizes on the UL and DL are same; however, the compression rate is different. It is observed that the UL has slightly better compression efficiency than DL in both scenarios, which is attributed to different compression methods used by UE and by eNB. In addition, the volume of information carried by the compressed data packets varies with the state in which the data packets are compressed. The decision about compression state transitions (from sending uncompressed data into compressing data with maximum compression efficiency) are made by the compressor based on many factors like the variations in the static part or dynamic part of packet headers and the acknowledgment feedback status from the de-compressor.

TABLE II
ROHC COMPRESSION EFFICIENCY FROM FIELD MEASUREMENTS

| AMR-NB with 12.65 Kbps and IPv4 | Concurrent VoLTE and Data sessions | | VoLTE session Only | | Grand Average |
|---|---|---|---|---|---|
| | UL | DL | UL | DL | |
| AMR Payload (Bytes) | 33 | 33 | 33 | 33 | 33 |
| Average Original Header Size (Bytes) | 40 | 40 | 40 | 40 | 40 |
| Other Protocol Overhead (L1/MAC/RLC/PDCP*) (Bytes) | 8 | 8 | 8 | 8 | 8 |
| Average Compressed Header Size (Bytes) | 3.9 | 7.5 | 3.2 | 6.5 | 5.3 |
| Average Compression Efficiency (%) | 90.1 | 81.2 | 91.9 | 83.6 | 86.7 |
| Required Channel Data Rate after ROHC (kbps) | 18.0 | 19.4 | 17.68 | 19.0 | 18.5 |

*L1: physical layer, MAC: medium access control layer, PDCP: packet data conversion protocol.

### 5.2 TTI Bundling

TTI bundling in general can achieve very good coverage and reliability [22]. TTI bundling optimization mainly depends on cell edge user RB utilization and SINR which are key factors to trigger TTI bundling, in fact the triggering criteria differ from one infra-vendor to another. However, careful optimization is required especially in the choice of the codec rate of VoLTE calls. For example, a VoLTE call with WB-AMR codec rate of 23.85 Kbps may face voice quality challenges even when TTI bundling is applied. In this scenario, every RTP packet sent on the UL will require an extra 8 ms to be transmitted. Assuming the maximum TBS of 63 bytes is granted to the UE during the TTI bundling operation (typical size during this operation) and when the ROHC is not used (either not configured or not applied due to ROHC state transition), the PDCP protocol data unit (PDU) size (including AMR payload and IPv4 RTP/UDP/IP headers) will be 102 bytes. In this scenario, one AMR payload cannot fit within one TTI and hence segmentation is needed. If ROHC is applied with the best compression efficiency (as demonstrated previously), the header size can be reduced to 3 bytes making the PDU arrives at the MAC layer with size of 65 bytes. In both cases, TTI bundling grant, to transmit a complete AMR payload, will not be sufficient over a single 4 msec bundle. Therefore, it requires more than one bundle to transmit one AMR payload (i.e. one RTP packet); hence, it can take ≥ 8 ms, which increases the delays, and in return deteriorate the voice quality at cell-edge. In all cases, it is obvious that the usage of ROHC and TTI bundling together at cell-edge engenders the optimum performance in terms of coverage and voice quality [21], [24], [27].

The lower codec rates provide larger coverage and better radio robustness because fewer voice information bits need to be sent over the air. Since media packets are generated in 20ms intervals then, the RTP Total Packet Size (RTPS) (i.e., one voice frame) can be estimated as follows:

$$RTPS = Payload + AMR\ header + RTP\ header + protocol\ header \quad (6)$$

Then the total RTP packet sizes for 12.65kbps and 23,85kbps codec rates can be estimated, respectively, as follows:

$$RTPS = 12.65x20ms + 11\ bits + 96\ bits + 64\ bits = 424\ bits \quad (7)$$

$$RTPS = 23.85x20ms + 11\ bits + 96\ bits + 64\ bits = 648\ bits \quad (8)$$

When TTI bundling is enabled, the resource allocation size is restricted to a maximum of three PRBs and the modulation scheme must be QPSK [28]. Therefore, the selected MCS index cannot be greater than 10. After TTI bundling is enabled, the maximum available TBS is as large as 504 bits that can be bundled in 4 TTIs (i.e., 504 LTE bits sent every 4ms). Therefore, one RTP packet of 424 bits can fit within the bundled TTI every 4ms for 12.65kbps while we need to send one RTP packet bundled every 8ms in case of 23.85kbps (i.e. two RTP packets bundled every 4 ms with the size of 648 bits). Since, VoLTE service is delay-sensitive, if higher-layer data is not transmitted within the specified delay budget, the voice quality deteriorates. This implies that RTP throughput is cut in half impacting the jitter and voice quality at cell edge, in addition to the negative impact on the uplink coverage. On the other hand, the TTI bundling does not apply to the DL, , instead methods like RLC segmentation concept are typically used to accommodate larger size RTP packets within smaller protocol layer packets at cell edge. Therefore, for codec rate like 23.85kbps, packets will be segmented into more smaller size RLC layer PDUs than 12.65kbps and hence higher voice delays can be observed on the downlink as well.

Based on the above discussion, it is recommended to reduce the codec rate to 12.65 Kbps to gain more uplink coverage with AMR payload that can fit within a bundled packet. This is another reason why the high definition codec rate of 23.85 kbps provides better voice quality in cell center while lower rates like 12.65 kbps provides better voice quality at cell-edge as demonstrated later in this section. The adaptive switching between different codec rates based on radio conditions or network load is still not widely applied. During an on-going call, the mobile-originated UE and mobile-terminated UE[*] have the ability to modify the codec mode and rate. The radio condition and user experience can be taken into account in this procedure. For the network load, the network congestion can be indicated by explicit congestion notification (ECN) mechanism via the RTP protocol using the RTCP as a feedback mechanism. However, ECN usage is very limited as it can only indicate the occurrence of congestion at the E-UTRA side without further information on its level. As for radio conditions indication between UE and eNB that can be used for codec rate adaptation, this mechanism is not clearly standardized at this point in 3GPP. Therefore, developing proper optimization processes is recommended to provide consistent user experience at all radio conditions. More details on the voice quality comparison between different codec rates are discussed later in this section. Also, the impact of TTI bundling on MOS and cell radius range are analyzed.

---

[*] Mobile-originated UE is the UE which originates the voice call while mobile-terminated UE is the UE which terminates the voice call.

### 5.3 RTP and Jitter Evaluation

We conducted a VoLTE to VoLTE long call in mobility conditions as summarized in table I and figure 4 and evaluated the performance of the KPIs mentioned in section IV. We have tested three different scenarios: concurrent VoLTE and Data connections with ROHC, VoLTE standalone call with ROHC and VoLTE standalone call with ROHC turned off from the eNB side. In all scenarios, the voice activity was continuous with minimum silent periods and we have performed full download using an FTP server when data session is present in first scenario.

The distributions (probability density function (PDF) and cumulative distribution function (CDF)) of the RTP error rate for all tested scenarios are illustrated in figure 5. With ROHC activated, the average downlink RTP error is within the accepted range of ≤ 1%. However, due to the presence of data packets in parallel with RTP voice packets, the RTP errors observed are slightly higher. This can happen in cases especially when both data and voice packets are multiplexed within the same TTI as explained in subsection (a) of this section. In this case, the HARQ scheduled with higher number of bits can jeopardize the VoLTE performance causing more air interface BLER. However, as will later be shown, the impact of higher BLER is more obvious on the jitter than the RTP error rate. On the other hand, when ROHC is disabled, the RTP error rate significantly degraded. This is because the VoLTE packets are transmitted with uncompressed headers leading to high packet sizes on the radio side. In return, this leads to more RTP errors and holes in IMS transmission (i.e. out of sequence) that impacts the overall voice quality. It is evident that enabling ROHC has positive impact on the performance alongside improvements to the capacity and coverage aspects as discussed previously.

From RTP jitter perspective, figure 6 illustrates the PDF and CDF of the downlink relative jitter in the same three scenarios. The presence of the concurrent PS data session significantly degraded the average relative jitter by 40% even with while ROHC is enabled. Additionally, the RTP error increased by 50% as shown in figure 5 however it is still within the accepted range. In the case of ROHC is disabled during VoLTE standalone call, the relative jitter is 20% higher compared to the same scenario with ROHC enabled. This stresses on the importance of ROHC feature to the VoLTE overall performance.

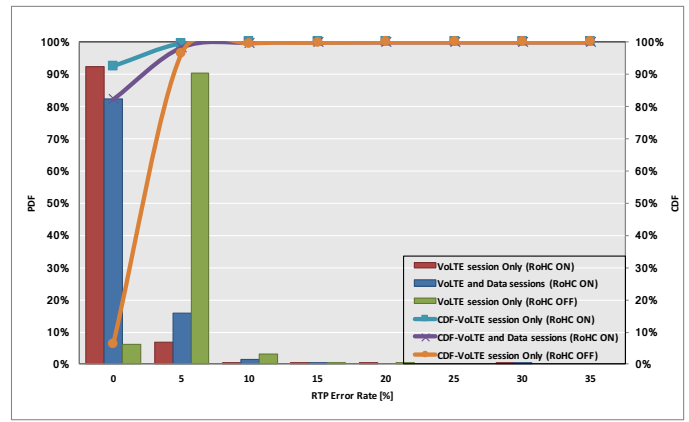

| Test Scenario for RTP Error rate | Avrg | Median | Min | Max |
|---|---|---|---|---|
| Concurrent VoLTE and Data sessions (ROHC ON) | 0.74% | 0% | 0% | 51.2% |
| VoLTE session Only (ROHC ON) | 0.34% | 0% | 0% | 54.2% |
| VoLTE session Only (ROHC OFF) | 7.77% | 7.84% | 0% | 34.7% |

Fig. 5. Distribution of the RTP error rate in mobility conditions

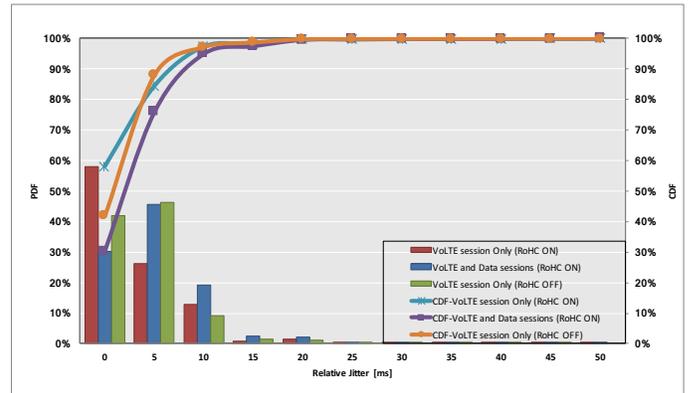

| Test Scenario for Relative Jitter | Avrg | Median | Min | Max |
|---|---|---|---|---|
| Concurrent VoLTE and Data sessions (ROHC ON) (ms) | 3.14 | 1.0 | 0 | 34.7 |
| VoLTE session Only (ROHC ON) (ms) | 1.87 | 1.5 | 0 | 25.6 |
| VoLTE session Only (ROHC OFF) (ms) | 2.32 | 1 | 0 | 220 |

Fig. 6. RTP DL jitter with ROHC enabled and without and with PS data

### 5.4 Scheduler Evaluation for VoLTE

The presence of PS data alongside VoLTE call has obvious degradation on the overall RTP performance as shown in figure 5 and Figure 6. It is therefore important to consider applying techniques to mitigate this negative impact. One of the options is to handle both data and VoLTE sessions in parallel at the eNB scheduler level. In this context, eNB scheduler can consider sending VoLTE packets in a given TTI without any data packets multiplexed especially in bad radio conditions. When the downlink scheduler tends to utilize the same physical layer transmission to send both voice and PS Data in the same TTI, then the TBS can increase and hence higher BLER. Another technique is to use the 2x2 MIMO codewords (or higher order MIMO) for splitting the VoLTE and PS data into two different streams with Rank-2 spatial multiplexing, i.e. segmentation of packets at MIMO codeword level. This can improve the spectral efficiency and minimize the jitter and overall BLER.

Table III provides scheduler comparison between the two infra-vendors used in this testing, for the case of concurrent

VoLTE and data sessions. The first scheduler does not tend to utilize the same Physical Downlink Shared Channel (PDSCH) transmission to send both IMS Media and PS Data in the same TTI. This means that more VoLTE media packets need to be transmitted alone which can increase waiting time in buffer and hence the overall jitter. Since multiplexing is not used, the TBS scheduled is low with high padded bits that waste the PRB/MCS resources and the overall capacity. Also, this scheduler does not seem to utilize MIMO Rank2 for IMS media packets and then there is no RLC segmentation over the two MIMO code-words. The second scheduler tends to utilize the same PDSCH transmission to send both IMS Media and PS Data in the same TTI. Since multiplexing is used in this deployment, the TBS scheduled is higher and the scheduler tends to minimize the padding and hence improve the PRB/MCS utilization and the overall capacity. Unlike the first scheduler, this one sometimes utilizes MIMO Rank 2 for IMS media packets and in this case use RLC segmentation over MIMO two code-words. This can add an additional improvement to the DL spectral efficiency.

TABLE III
SCHEDULER IMPLEMENTATION COMPARISON

| Scheduler Behavior | Scheduler 1 | Scheduler 2 |
|---|---|---|
| IMS Media Packets sent without Data (i.e., packets are not multiplexed) | 88% | 53% |
| IMS Media + PS Data multiplexed in same TTI | 12% | 47% |
| IMS packets transmitted using MIMO two code-words | 0% | 5.42% |
| Inter-TTI scheduling delay (i.e., delay between two MAC layer packets scheduled by eNB carrying IMS media) (ms) | 34.3 | 20.9 |
| Average TBS Scheduled (Bytes) | 585 | 1291 |
| Average padding per TBS at MAC (Bytes) | 302 | 41 |

So far, we have analyzed the relative jitter in different scenarios with concurrent PS data, during VoLTE standalone call and also with and without ROHC. In this section, we will analyze the relative jitter versus BLER at the radio interface. Figure 7 demonstrates the relative jitter versus DL BLER with and without data session. As evident from figure 7, there is a clear relation between relative downlink jitter and downlink BLER at physical layer as there is trend of jitter increase with the increase of re-transmission at lower layers. It is also obvious that there is high variation in jitter when PS data is present. More specifically, the jitter reached an average of 3.5 ms at BLER of 10% with PS data while the jitter reached an average of 2 ms at the same BLER level and with VoLTE standalone call. This implies that DL BLER and the BLER target at eNB can affect the end-to-end RTP delays and therefore careful implementation of BLER convergence algorithms at the eNB scheduler is mandatory.

Due to delays at different interfaces, the RTP packets transfer with different inter-arrival intervals. From eNB side, the convergence into the targeted BLER requires a stable flow of packets in order to maintain a suitable tracking time to achieve the target. Typically, the eNB scheduler selects the downlink TBS based on the reported CQI which controls the MCS selection. The amount of data to be scheduled in a TTI determines the number of PRBs to be scheduled for UEs. Based on the feedback mechanism between UE and eNB, the scheduler keeps tracking the BLER measured versus the target BLER and hence starts the adjustment of the MCS and PRB to this UE. This mechanism enables the scheduler to allocate resources in a manner to maximize the resources utilization efficiency. However, in some cases where the packets flow irregularly, the tracking mechanism cannot converge which can cause fluctuations of BLER as observed in figure 7. To maintain a good tradeoff between DL BLER and DL resources, the scheduler should be designed to minimize the TBS (and the padding bits in each TBS) while maintaining a spectral efficiency in terms of good throughput.

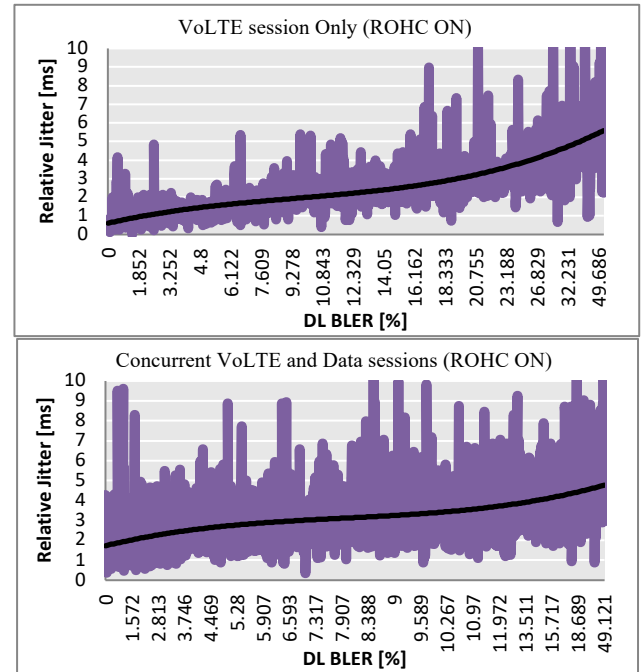

Fig. 7. RTP relative jitter *vs.* DL BLER

TABLE IV
DL SCHEDULING AND THROUGHPUT COMPARISON BETWEEN VoLTE AND OTT

| Near Cell Conditions (VoLTE scenarios operating at 12.65 Kbps and IPv4) | OTT | VoLTE (ROHC OFF) | VoLTE (ROHC ON) |
|---|---|---|---|
| Average Time Domain Scheduling Rate (%) | 6% | 2% | 2% |
| Average MAC Layer TBS (bytes) | 360.8 | 344.4 | 163.1 |
| Average Scheduled Resource Blocks | 4.2 | 5.4 | 4.1 |
| Average Bit Rate (kbps) | 155.1 | 54.7 | 25.4 |

To evaluate the scheduler behavior, we compare the downlink throughput of three cases: VoLTE with and without ROHC and Over-the-Top (OTT) VoIP using a commercial application. Typical OTT applications include Skype or Facetime, where the IP packet transmission is handled without the network management of an operator. In this case, the OTT utilize QCI=9 as a normal PS data session carrying VoIP packets. Table IV provides a comparison between VoLTE (with and without ROHC) and OTT in terms of radio resource utilization at different layers based on scheduling attempts. As shown in table IV, OTT consumes the most scheduling resources in the time domain which means it would require more TTI utilization compared to VoLTE in general, while the PRB utilization in the frequency domain is almost similar for VoLTE and OTT. As a result, OTT will require higher throughput to maintain the voice quality and that consumes more scheduling resources from the network, affecting the overall cell capacity. As shown, OTT requires six times more data throughput compared to VoLTE with ROHC, which means more resources and mainly consumed in time-domain. It is worth mentioning that this result is for the tested networks and with one OTT application. On the other hand, VoLTE with ROHC enabled provides the most efficient scheduling and the lowest data rate. This indicates that when VoLTE deployed with optimized feature will outperform OTT in cell capacity aspects. However, it is to be noted that the TBS in all cases fairly exceeds the actual RTP payload (i.e. 72 bytes without ROHC and ~ 37 bytes with ROHC). This indicates that the scheduler still strives to meet proper scheduling size when the IMS packets transfer irregularly. This is also an observation that may require further investigation in the scheduler behavior and in future work in VoLTE evolution. The indication of the TBS and the padding bits are highlighted in table III. With this data, we observe that network schedulers have a room of improvements to map PRB to MCS in a way to utilize sufficient TBS and minimize the padding bits and consequently the LTE spectrum efficiently.

## 5.5 RTP and Jitter Evaluation versus Radio Conditions

In this section, we evaluate the RTP performance versus radio conditions. Figures 8 to 10 demonstrate the average RTP error rate and jitter as a function of RSRP, RSRQ and SINR, respectively. The figures illustrate that the jitter and RTP error rate tend to increase when the RF conditions degrade. However, the most impacting factor is the RSRQ (loading and interference indicator) where the jitter and RTP error rate exhibit significant degradation (i.e., jitter > 10 ms and RTP error rate > 10%) when the RSRQ is degraded. This indicates that high interference or high loading at the eNB will directly affect the scheduling of the RTP packets and thus increases the RTP delays. It is therefore important to optimize the handover parameters based on RSRQ as well as RSRP to improve the overall radio conditions of VoLTE performance. RSRQ is a typical quality measure of the loading and interference in connected mode. In a multiple bands scenario, the RSRQ can be used as a trigger to distribute the load uniformly across the two bands. The RSRP reflects only the channel condition, e.g. if UE is near or far from cell center, however the RSRQ can indicate whether the cell on that carrier is loaded or not. Hence, the RSRQ can be used to trigger an inter-frequency handover from loaded carrier into an unloaded one. As the voice service is more sensitive to the radio and loading conditions, then improving the handover trigger based on RSRQ can be beneficial to VoLTE calls. Moreover, The eNBs can be configured with different handover parameters for VoLTE call compared to other data sessions. This can be configured since the eNB is aware of the VoLTE and data bearers and then it can define the handover parameters differently for each service. More importantly, the jitter starts to increase significantly when the RSRP < -110 dBm. This can be a decisive factor to set the threshold for triggering eSRVCC to 3G or 2G to maintain a good voice quality. With this recommended setting, the coverage of VoLTE service will reach up to RSRP of -110 dBm, beyond which, the UE will be handed over to 3G/2G using eSRVCC in order to enhance voice quality.

## 5.6 Handover Impact on VoLTE Call Performance

It is expected that users during VoLTE call will be in mobility conditions where handovers between different eNBs will occur frequently. Therefore, evaluating the delays of RTP packets during the VoLTE call is important to ensure consistent voice quality in mobility conditions. The authors in [20] evaluated the performance of VoLTE call in mobility conditions for both intra and inter-systems handovers. The results showed that inter-system handovers incurred higher delays than intra-system handovers. They evaluated both the user and control plane delays associated with these handover types. In this paper, we will evaluate the impact of intra-frequency handover on the characteristics of VoIP call represented by the jitter and RTP errors that could occur during mobility conditions. Figure 11 demonstrates the intra-frequency handover procedures and the associated delays affecting the RTP interruption and jitter.

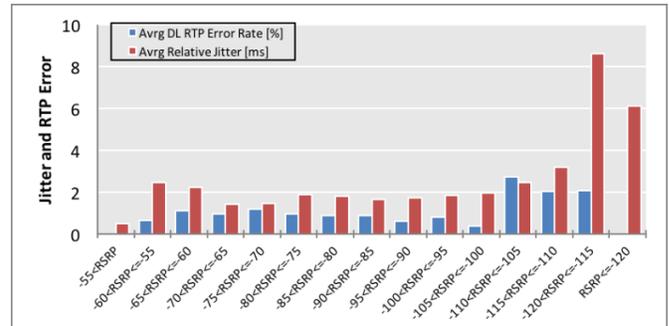
Fig. 8. average RTP and Jitter vs. RSRP

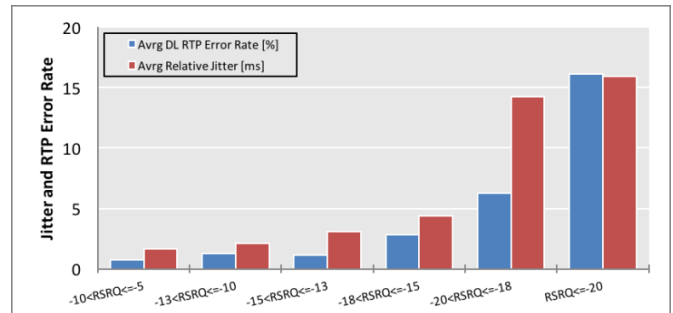
Fig. 9. average RTP and Jitter vs. RSRQ

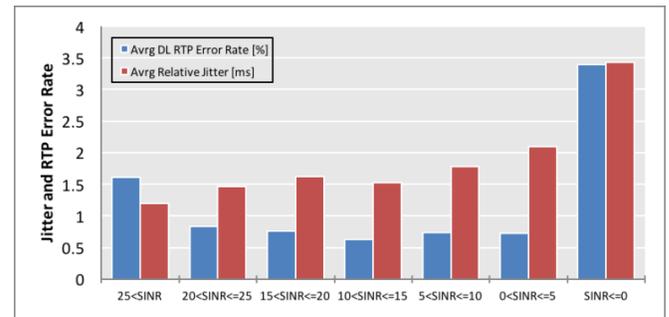
Fig. 10. average RTP and Jitter vs. SINR

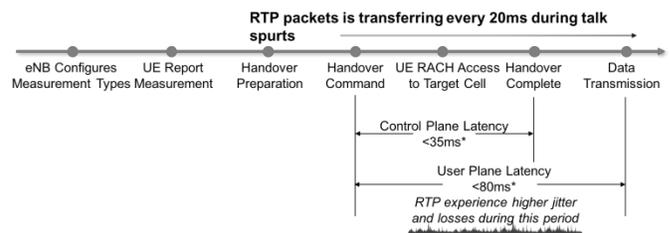
* refer to [20] for more details
Fig. 11. LTE Intra-Frequency Handover Procedures

As explained in figures 8 to 10, the radio conditions have direct impact on the VoLTE performance. We will go one-step further by evaluating the jitter and interruption during the handover procedure. Handover is essential during mobility to move the user from one eNB to another. If the parameters are set to delay the handover, then higher jitter and errors can be observed as discussed before in terms of

RSRP and RSRQ. The analysis in [20] presents the details of the voice interruption time where the actual RTP packets are suspended during the handover execution. We here evaluate the distribution of the RTP interruption and jitter during the handover execution procedure as shown in figure 12, with ~100 handover attempts in the testing route. As observed from figure 12, the average jitter during handover is higher than the normal average observed in figure 6. As estimated in [20], the average interruption time is ~ 75 ms, which means that the jitter is directly impacted by the delay in transferring the VoLTE context from one eNB to another over X2 interface. Assuming that the RTP packets are not lost as evident from the RTP error rate in figure 12, then the time difference between the jitter and RTP interruption is ~ 40 ms. This means that there is ~ 40 ms delay in downlink scheduling of two consecutive RTP packets during handover. The reduction of such interruption time is very critical especially for certain applications that utilize VoIP services. This topic requires further research for enhancements and it can be proposed as contribution in LTE-Advanced Pro (3GPP Rel-13 and beyond).

It is important to know that the IMS server and client maintain timers to detect any timeout in RTP packet transfer, after which the VoLTE call can be released. In this context, UE IMS client detects that the DL RTP packets are not detected within a certain time window, e.g. for 10 sec, then the IMS client will terminate the voice call to prevent the call from being active without audio. Therefore, optimizing the handover performance is essential to keep the established VoLTE with good retainability and consistent performance.

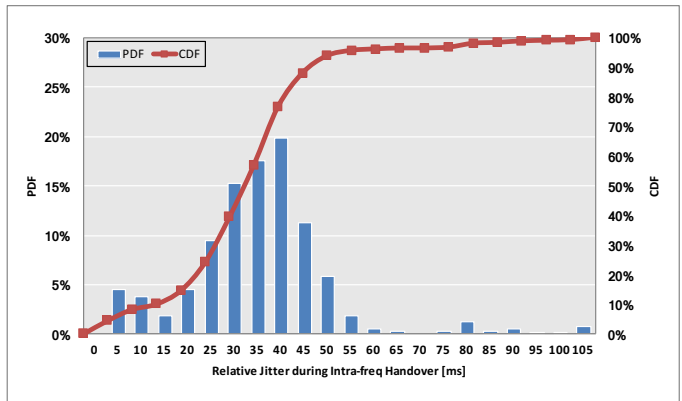

| KPIS | Avrg | Min | Max |
|---|---|---|---|
| Number of RTP Packet lost during handover | 0.33 | 0 | 21 |
| Relative DL Jitter during handover (ms) | 35.6 | 2 | 922 |

Fig. 12. RTP performance during Intra-Frequency Handover execution

### 5.7 VoLTE Voice Quality Evaluation

In this section, we will evaluate the VoLTE voice quality in terms of POLQA (Perceptual Objective Listening Quality Analysis) MOS score [29] and compare it with OTT application (as described earlier in section), 3G with AMR-WB and 2G with AMR-NB. POLQA was adopted in 2011 as ITU-T Recommendation P.863 [30]. In this testing, we have used same clusters as described in section IV. The voice quality testing is conducted based on mobility with average speed of 80km/h. We used Spirent Nomad voice quality tool with 4 channels for testing. In each test scenario, we used four devices as two for UL and two for DL. In the first time, we conducted 2G and 3G and in the second time we used two VoLTE calls with different coding rate i.e., 23.85kbps and 12.65kbps. The test devices are locked on the tested network i.e., 2G, 3G or LTE with no handover or SRVCC. The OTT application is tested separately and the device was locked on LTE. The calls are long calls i.e, around 16 min which is around 50 cycles of the MOS testing device. We averaged the DL and UL and averaged the results over 10 times for each round.

Figure 13 provides the average MOS values for VoLTE with codec of 23.85, VoLTE with codec 12.65, OTT, 3G with AMR-WB of 12.65 and 2G AMR-NB. The figure indicates that VoLTE engenders the best voice quality compared to OTT and CS voice calls. Also, the average MOS of VoLTE with 12.65 kbps codec rate is better than VoLTE with 23.85 kbps codec rate. This is because the 12.65 kbps is more robust at cell-edge. Therefore, in mobility scenario and since MOS values are averaged, the overall measured MOS of VoLTE with 12.65 kbps is better than VoLTE with 23.85kbps. It is also noted that, LTE with 12.65 Kbps falls within the range of "good quality" which is specified in ITU-T P.863 as $3.6 \leq MOS \leq 4.0$. On the other side, 3G and OTT fall in the range of "acceptable quality" specified as $3.1 \leq MOS \leq 3.6$, while the 2G falls into the "poor quality" of $2.6 \leq MOS \leq 3.1$, as it is not high definition.

Figure 14 provides comparison between VoLTE with the two codec rates at near cell and far cell scenarios. The RSRP range for near cell scenario is -60 to -65 dBm and the far cell scenario is -102 to -105 dBm. It is evident that the VoLTE with 23.85kbps codec offers bad voice quality at cell-edge as explained earlier in TTI bundling section. Therefore, it is recommended to use the speech rate of 12.65kbps to guarantee a consistent user experience. Also, 23.85kbps would consume higher resources compared to 12.65kbps (due to higher payload size) however with a minimal improvement to the voice quality in near cell and with highly degraded quality at the cell-edge. EVS codec was not evaluated in this paper but it is anticipated that EVS with SWB can outperform AMR-WB even at mixed and music content and offers 1.2 MOS improvement at comparable bitrates [31].

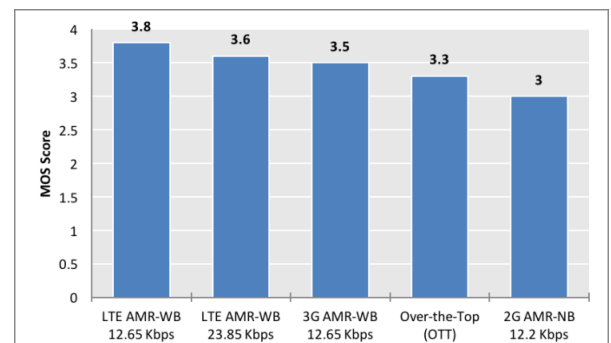

Fig. 13. Voice Quality Measurements for Different Technologies (Mobility)

Finally, in order to evaluate the TTI bundling gain at cell-edge, we have conducted a field test to measure the average MOS at cell-edge with and without TTI bundling. Figure 15 provides the average MOS versus RSRP for the two scenarios. As depicted form the figure, the TTI bundling improves MOS when RSRP≤ -118dBm. This is the threshold that can be set to trigger the TTI bundling. Forcing the TTI bundling at RSRP >-118dB unnecessarily repeats the packet four times and hence increasing the jitter and reducing the MOS. The TTI bundling extends the cell-edge to RSRP = -129dBm while without TTI bundling, the cell radius is limited to RSRP =~ -123dBm (i.e., call dropped at this level).

Therefore, the TTI bundling has extended the coverage by 6dB. This gain can be beneficial for new technologies such as 3GPP Narrow Band Internet of Things (NB-IoT) that relies on the extended coverage concept and requires additional 20dB increase in LTE coverage to reach underground and deep indoor sensors [32], [33].

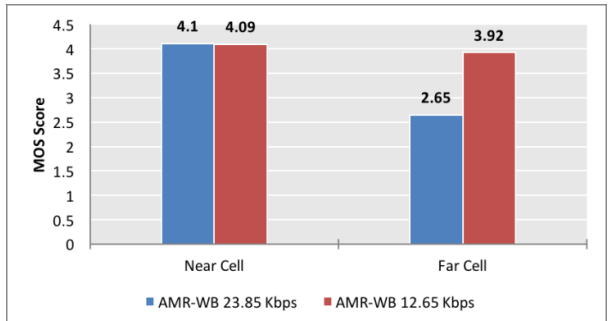

Fig. 14. Voice quality for different VoLTE codec rates at different RF conditions

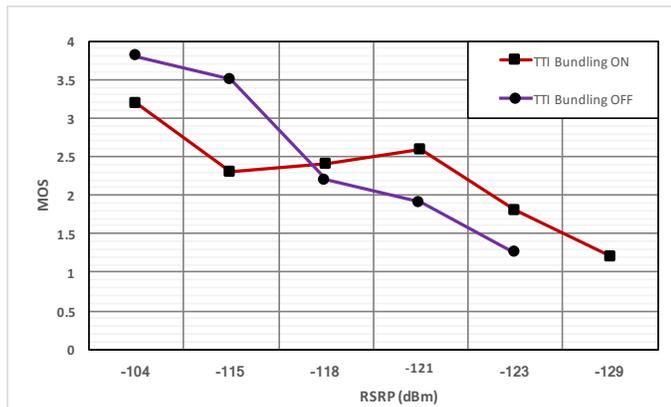

Fig. 15. TTI Bundling impact on MOS

## 8 Conclusions

In this paper, we have analyzed the practical performance of VoLTE based on commercially deployed 3GPP Release-10 LTE networks. The evaluation demonstrates VoLTE performance in terms of ROHC, RTP error rate, RTP jitter and delays, TTI bundling, BLER and VoLTE voice quality in terms of MOS. This paper provided best deployment practices for VoLTE deployment and VoLTE related features. We have demonstrated that ROHC is capable of offering significant gain to the radio resources by reducing the packet size and compressing the headers with an average efficiency of 81% to 92% and overall average efficiency of 86.7% based on evaluated networks. Accordingly, ROHC can boost the air interface resources by almost twofold. Also, ROHC offers ~7% improvement in RTP error rate and ~24% reduction in jitter. The use of ROHC and TTI bundling together at cell-edge will be the best in terms of improved coverage and best voice quality. However, it is recommended to reduce the codec rate to 12.65 Kbps to gain more uplink coverage with AMR payload that can fit within a bundled packet.

The concurrent data and VoLTE session causes 40% degradation in jitter and 50% increase in the RTP error however they are still within the accepted ranges. Also, the jitter reached an average of 3.5 ms at BLER of 10% with PS data while the jitter reached an average of 2 ms at the same BLER level with VoLTE standalone. The presence of PS data alongside VoLTE call has obvious impact to the overall RTP performance. We have provided techniques to mitigate these drawbacks. The degradation in the RSRQ leads to significant degradation in the jitter and RTP error rate. More importantly, the jitter starts to increase significantly when the RSRP < -110 dBm. Therefore, it is recommended to limit the coverage of VoLTE service to RSRP =-110dBm by trigging SRVCC. However, TTI bundling can extend VoLTE coverage to RSRP = -129dBm. There is ~ 40 ms delay in downlink scheduling of two consecutive RTP packets during handover. The reduction of such interruption time is very critical especially for certain applications that utilize VoIP services.

we have evaluated VoLTE in terms of voice quality using POLQA MOS. It is demonstrated that VoLTE engenders the best voice quality compared to CS voice calls. Also, the average MOS of VoLTE with 12.65kbps is better than VoLTE with 23.85kbps. It is highlighted that 23.85kbps offers very bad voice quality at cell-edge. Therefore, it is recommended to fix the codec rate at 12.65kbps to guarantee consistent user experience. Future work may include proposing techniques to reduce the RTP data interruption of 40msec during handover. Also, evaluating the adaptive codec selection to guarantee best voice quality and minimize radio resources utilization.

## References

[1] 3GPP TS 22.228 V12.9.0, Service requirements for the Internet Protocol (IP) Multimedia core network Subsystem (IMS).

[2] 3GPP TS 22.173 V9.5.0 (2010-03): "IP Multimedia Core Network Subsystem (IMS) Multimedia Telephony Service and supplementary services; Stage 1 (Release 9)"

[3] GSMA, IR.92 IMS Profile for Voice and SMS, Version 7.0, 03 March 2013.

[4] GSMA, IR.94 IMS Profile for Conversational Video Service Version 5.0 04 March 2013.

[5] A. Elnashar, M. A. El-Saidny, "Looking at LTE in Practice: A Performance Analysis of the LTE System based on Field Test Results," IEEE Vehicular Technology Magazine, Vol. 8, Issue 3, pp. 81:92, Sept. 2013.

[6] Ayman Elnashar, Mohamed Al-saidny, and Mahmoud Sherif "Design, Deployment, and Performance of 4G-LTE Networks: A practical Approach," Wiley, May 2014.

[7] 3GPP TS 26.441, "Codec for Enhanced Voice Services (EVS); General overview", version 12.0.0 Release 12

[8] Myasar R. Tabany and Chris G. Guy, "Performance Analysis and Deployment of VoLTE Mechanisms over 3GPP LTE-based Networks," International Journal of Computer Science and Telecommunications, Volume 4, Issue 10, October 2013.

[9] Myasar R. Tabany, Chris G. Guy, An End-to-End QoS Performance Evaluation of VoLTE in 4G E-UTRAN-based Wireless Networks, ICWMC 2014, pp. 90:97.

[10] J. Calle-Sánchez, M. Molina-García, J. I. Alonso, and A. FernándezDurán, "Long term evolution in high speed railway environments: feasibility and challenges," Bell Labs Tech. J., vol. 18, no. 2, pp. 237– 253, 2013.

[11] Sniady, A.; Sonderskov, M.; Soler, J., VoLTE Performance in Railway Scenarios: Investigating VoLTE as a Viable Replacement for GSM-R, IEEE Vehicular Technology Magazine, 2015, Volume: 10, Issue: 3, Pages: 60 - 70

[12] J. Seo and V. Leung , "Performance modeling and stability of semi-persistent scheduling with initial random access in LTE", IEEE Trans. Wireless Commun. , vol. 11 , no. 12 , pp.4446 -4456 , 2012

[13] Q. Bi, S. Vitebsky, Y. Yang, Y. Yuan, and Q. Zhang, "Performance

[14] J. A. Fernández-Segovia, S. Luna-Ramírez, M. Toril, and Juan J. Sánchez-Sánchez, IEEE Communications Letters, VOL. 19, NO. 3, MARCH 2015 431 Estimating Cell Capacity From Network Measurements in a Multi-Service LTE System

[15] A. Sanchez-Esguevillas, B. Carro, G. Camarillo, Y. -B. Lin, M. A. Garcia-Martin and L. Hanzo , "IMS: the new generation of Internet-protocol-based multimedia services" , Proc. IEEE , vol. 101 , no. 8 , pp.1860 -1881

[16] J. E. Vargas Bautista et. all, "Performance of CS Fallback from LTE to UMTS" IEEE Communications Magazine, pp. 136:143, Sept. 2013.

[17] R.-H. Liou et. all, "Performance of CS Fallback for Long Term Evolution Mobile Network," IEEE Transactions On Vehicular Technology, VOL. 63, NO. 8, pp. 3977:3984, OCTOBER 2014.

[18] Yi-Bing Lin, "Performance Evaluation of LTE eSRVCC with Limited Access Transfers," IEEE Transactions On Wireless Communications, VOL. 13, NO. 5, MAY 2014.

[19] 3GPP TS 23.272 V10.3.1 (2011-04), Circuit Switched (CS) fallback in Evolved Packet System (EPS).

[20] Ayman Elnashar, Mohamed A. El-Saidny, Mohamed Mahmoud, "Practical Performance Analyses of Circuit Switched Fallback (CSFB) and Voice over LTE (VoLTE)," IEEE Transactions on Vehicular Technology, Vol. 66, Issue 2, pp. 1748 - 1759

[21] Andreas Maeder and Arne Felber, "Performance Evaluation of ROHC Reliable and Optimistic Mode for Voice over LTE," In Proc. Vehicular Technology Conference (VTC Spring), 2013 IEEE 77th.

[22] M. Villaluz et. all, VoLTE SRVCC Optimization as Interim Solution for LTE Networks with Coverage Discontinuity, International Conference on Information and Communication Technology Convergence (ICTC), 2015 , pp. 212:216.

[23] 3GPP TS 26.114, IP Multimedia Subsystem (IMS); Multimedia telephony; Media handling and interaction

[24] Daniel Philip VENMANI et. all, "Impacts of IPv6 on Robust Header Compression in LTE Mobile Networks," ICNS 2012: The Eighth International Conference on Networking and Services, pp. 175:180.

[25] NSN white paper, "From voice over IP to voice over LTE" Nov. 2013.

[26] 3GPP TS 36.321, Medium Access Control (MAC) protocol specification, version 12.5.0 Release 12

[27] O. Ozturk and M. Vajapeyam, "Performance of VoLTE and data traffic in LTE heterogeneous networks", IEEE GLOBECOM, 2013.

[28] 3GPP TS 36.213, Evolved Universal Terrestrial Radio Access (E-UTRA); Physical layer procedures, version 12.4.0 Release 12

[29] ETSI CTI Plug tests Report, VoLTE QoS Assessment, 0.1.0 (2013-12)

[30] ITU-T POLQA Recommendation P.863, website http://www.itu.int/rec/T-REC-P.863/en

[31] Fraunhofer Institute for Integrated Circuits IIS, Enhanced Voice Services (EVS) Codec, Technical paper: (online) http://www.iis.fraunhofer.de/content/dam/iis/de/doc/ame/wp/FraunhoferIIS_Technical-Paper_EVS.pdf

[32] TR 45.820 v13.1.0, "Cellular system support for ultra low complexity and low throughput internet of things," Nov. 2015.

[33] A. Elnashar, Mohamed A. El-Saidny, "Practical Guide to LTE-A, VoLTE and IoT: Paving the way towards 5G," Wiley, July 2018.

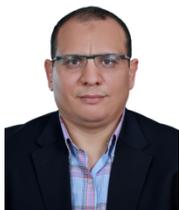

**Ayman Elnashar** received the B.S. degree in electrical engineering from Alexandria University, Egypt, in 1995 and the M.Sc. and Ph.D. degrees in electrical communications engineering from Mansoura University, Egypt, in 1999 and 2005, respectively. He has 20+ years of practical experience and leadership positions in ICT industry including planning, deployment and operation. He was part of three major start-up service providers (Orange/Egypt, Mobily/KSA, and du/UAE). Currently, he is VP and Head of Infrastructure Planning – ICT & Cloud with the Emirates Integrated Telecommunications Co. "du", UAE. He is the founder of the Terminal Innovation Lab and UAE 5G innovation Gate (www.u5gig.ae) to derive 5G adoption in UAE. He is leading the mobile/fixed network transformation towards SDN/NFV and digital transformation of EITC Infrastructure to become a digital ICT service provider. Prior to this, he was Sr. Director – Wireless Networks, Terminals and IoT where he managed and directed the evolution, evaluation, and introduction of du wireless networks, terminals and IoT including LTE/LTE-A, HSPA+, WiFi, NB-IoT and currently working towards deploying 5G network in UAE. Prior to this, he was with Mobily, Saudi Arabia, from June 2005 to Jan 2008 as Head of Projects. He played key role in contributing to the success of the mobile broadband network of Mobily/KSA. From March 2000 to June 2005, he was with orange Egypt.

He published 30+ papers in wireless communications and wireless networks in highly ranked journals and international conferences. He is the author three books published by Wiley:

- Design, Deployment, and Performance of 4G-LTE Networks: A Practical Approach, published in May 2014,
- Practical Guide to LTE-A, VoLTE and IoT: Paving the Way Towards 5G, published in July 2018,
- Simplified Robust Adaptive Detection and Beamforming for Wireless Communications, published in July 2018.

His research interests include practical performance analysis, planning and optimization of wireless networks (3G/4G/WiFi/IoT/5G), digital signal processing for wireless communications, multiuser detection, smart antennas, massive MIMO, and robust adaptive detection and beamforming.

He is more focused now on network transformation towards SDN/NFV, Cloud transformation, digital transformation, ICT applications, IoT evolution and 5G use cases.

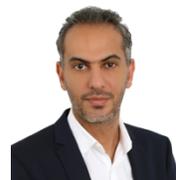

**Mohamed A. El-saidny** received the B.Sc. degree in Computer Engineering and the M.Sc. degree in Electrical Engineering from the University of Alabama in Huntsville, USA in 2002 and 2004, respectively. He is currently Director - Technology at MediaTek. He is a leading technical expert in wireless communication systems for modem chipsets and network design. He is driving a team responsible for the technology evolution and the alignment of the network operators to the device and chipset roadmaps/products in 3G, 4G and 5G. His main focus is on expanding MediaTek technologies and technical expertise with the mobile network operators worldwide. Prior to MediaTek, he worked at Qualcomm CDMA Technology, Inc. (QCT), San Diego, California, USA. He later moved to mobile network design in Qualcomm's Corporate Engineering Services division in Dubai, UAE. At Qualcomm, he was responsible for performance evaluation and analysis of the UMTS and LTE system solutions for user equipment. He developed and implemented system studies to optimize the performance of various UMTS and LTE algorithms including Cell re-selection, Handover, Cell Search and Paging, CSFB, C-DRX, Inter-RAT, VoLTE/IMS, Carrier Aggregation and multi-band load balancing techniques. His current research interest is on the 5G evolution and gap analysis in 5G requirements compared to the 4G deployment challenges in the areas of physical layer, High Reliable/Low Latency systems, and waveforms design concepts. He is the inventor of numerous patents in CDMA and FDMA systems, the co-author of "Design, Deployment and Performance of 4G-LTE Networks: A Practical Approach" book by Wiley & Sons, in addition to contributions to 3GPP algorithms. He published several international research papers in IEEE Communications Magazine, IEEE Vehicular Technology Magazine, and other IEEE Transactions.

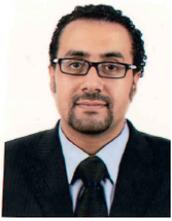 **Mohamed Yehia** received the B.S. degree in electrical communications engineering from Cairo University, Egypt, in 2006. Currently, he is currently Manager of Terminals and Performance with the Emirates Integrated Telecommunications Co. "du", UAE. He is responsible for testing and validation of new terminals/chipsets and new 2G/3G/LTE features. He has over all 9 years of experience in multi-culture environments, focusing on GSM, WCDMA and LTE protocols, troubleshooting, features testing and deployment, capacity planning, network dimensioning, network optimization and network performance monitoring. And Having management and technical experience within different projects for network implementations, managed services in Middle East and Africa, he has proven experience in Wireless 2G/3G/HSPA/HSPA+/LTE/LTE-A.